\documentclass[prx,superscriptaddress,twocolumn,showpacs,preprintnumbers,amsmath,floatfix]{revtex4-1}
\usepackage{amssymb}
\usepackage{graphicx}

\begin{document}

\title{Magnetism and Superconductivity in Iron-based Superconductors Decided by Condensed Particle-hole Excitations away from the Fermi Level}
\author{Ming-Cui Ding}
\affiliation{Shanghai Key Laboratory of Special Artificial Microstructure Materials and Technology, School of Physics Science and engineering, Tongji University, Shanghai 200092, China}
\author{Hai-Qing Lin}
\affiliation{Beijing Computational Science Research Center, Beijing 100084, China}
\author{Yu-Zhong Zhang}
\email[Corresponding author.]{Email: yzzhang@tongji.edu.cn}
\affiliation{Shanghai Key Laboratory of Special Artificial Microstructure Materials and Technology, School of Physics Science and engineering, Tongji University, Shanghai 200092, China}
\affiliation{Beijing Computational Science Research Center, Beijing 100084, China}
\date{\today }

\begin{abstract}
The origin of magnetism and superconductivity in iron-based superconductors is still unclear. Here, by investigating the momentum-dependent particle-hole excitations which quantify the tendency of itinerant electrons towards various magnetic states or superconducting phases, we unravel a novel origin to account for the variety of physical properties of iron-based compounds. We show that condensation of particle-hole excitations away from the Fermi surface in momentum space is the underlying mechanism in deciding the magnetic and superconducting properties of iron-based materials. The applicability of this scenario to the whole family of iron-based superconductors suggests that inclusion of the orbital degrees of freedom, which may lead to competing tendencies towards different magnetically ordered states, is more crucial than taking into account the strong correlations. Our findings further indicate that in order to properly model these materials, the electronic states away from the Fermi level have to be considered.
\end{abstract}
\pacs{71.20.-b,74.70.Xa,75.10.Lp,71.10.Fd}

\maketitle

\section{Introduction}

High temperature superconductivity always attracts scientists working in different fields of physics and material science due to its extensive applications and extraordinary properties. A fundamental question of this topic is how to understand the origin of high temperature superconductivity~\cite{Norman}? Although preliminary consensus has been reached with respect to the high-T$_c$ cuprates that strong correlation is indispensable for high temperature superconductivity~\cite{LeeNagaosaWen}, intensive debates persist since the discovery of high-T$_c$ iron-based superconductors.

In contrast to the strongly-correlated high-T$_c$ cuprates, magnetism and superconductivity in iron-based superconductors was originally proposed to be understandable from both the strong-coupling limit, where spin exchange effect of localized electrons is emphasized, and weak-coupling limit, where condensation of particle-hole excitations among the Fermi surfaces is highlighted~\cite{Wanglee,Paglione,Mazin_PhysicaC,Pickett,Mazin-Singh}. However, while conventional weak-coupling theory of the Fermi surface nesting~\cite{Eremin,bkfeas} is continuously being challenged as more experiments are performed~\cite{BaoFeTe,He-NaFeAs,DaiHu}, the density functional theory calculations in combination with dynamical mean field theory do not support the existence of localized electrons in Fe $3d$ orbitals~\cite{YinHauleKotliar}.

Recently, a compromised scenario for understanding the iron-based superconductors was proposed, where localized and itinerant electrons are assumed to coexist in Fe $3d$ orbitals~\cite{KouLiWeng,HacklVojta,YinLeeKu}. Though, angular resolved photoemission spectroscopic study on A$_x$Fe$_{2-y}$Se$_2$ ($A$=K, Rb), corroborated by a slave-spin mean-field calculation, pointed to a possible observation of such a coexistence at finite temperature~\cite{OSMPexp},  it is still under debate experimentally whether the studied sample or which part of the studied sample is the parent compound of the superconducting state~\cite{Daggoto}. In other words, the detected coexistence may not be relevant to superconductivity. 

Another origin for magnetism in iron-based superconductors was proposed based on comparisons of densities of states between nonmagnetic state and various magnetic states in FeTe and BaFe$_2$As$_2$ obtained from density functional theory calculations. It was concluded that the magnetic ground state is determined by all occupied states below the Fermi level, not by fermiology~\cite{JohannesMazin}. However, on one hand, strong downshift in spectral weight due to spin polarization usually happens in all the materials with large magnetic moment. On the other hand, as is well known, the magnetic moments in most iron-based compounds are overestimated by density functional theory calculations~\cite{OpahleZhang,ZhangOpahle}. Therefore, this scenario is also questionable.

In this paper, by applying density functional theory calculations to various families of iron-based superconductors, we proposed that the condensed particle-hole excitations away from the Fermi level, rather than those among the Fermi surfaces, are responsible for the distinct physical properties among different iron-based compounds. We further pointed out that the orbital degrees of freedom, inter-atomic magnetic interaction, and interlayer couplings have to be involved in order to correctly understand some anomalies in iron-based compounds. We found that while the strong correlation is not indispensable if one wants to understand magnetism and superconductivity of iron-based materials at qualitative level, the electronic states above and below Fermi level are both important.

\section{Computational Details}

\subsection{Density Functional Theory Calculations}
We use experimentally-determined structures for all the compounds (see appendix~\ref{sec:app-one}), i.e.~the [1111] family LnOFePn with Ln=Sm, Nd, Ce, La and Pn=As, P through to SrFFeAs, the [111] family containing AFePn with A=Li, Na, the [11] family FeCh with Ch=Se and Te, and the [122] family including AeFe$_2$Pn$_2$ with Ae=Ca, Sr, Ba. We temporarily leave KFe$_2$Se$_2$ to later study because the parent compound of its superconducting state is still under debate experimentally and the observed special magnetic states are always connected to different types of iron vacancies~\cite{Daggoto}. Both the paramagnetic and spin-polarized cases are calculated by the full-potential linearized augmented plane wave method as implemented in Wien2k~\cite{balaha} with the exchange correlation functional of Perdew, Burke and Ernzerho (PBE). The results are consistent with those calculated within the local density approximation (LDA). We choose RK$_{max}$=7 and total 40000~k points in the Brillouin-zone integration. The open core approximation is employed when the f electrons are treated in the nonmagnetic states. In the LDA+U or the spin polarized GGA+U calculations, f electrons of Ce atoms are arranged ferromagnetically according to the experiments, and $U$ is chosen to be 6~eV~\cite{Ceatom}. Moreover, different $U$ leads to qualitatively the same results. Here the atomic limit double-counting correction is taken into account since f electrons on Ce atoms are strongly correlated~\cite{balaha}.

\subsection{momentum-dependent particle-hole excitations within constant matrix elements approximation}
\label{sec:one}

The condensation of the momentum-dependent particle-hole excitations can be quantified by the real part of the static bare susceptibility. Within constant matrix elements approximation~\cite{Mazin-Singh}, the static bare susceptibility reads:
\begin{eqnarray}
  \chi_{0} (q)&=& -\frac{1}{N} \sum_{k,\mu\nu}\frac{1}{E_{\nu}(k+q)-E_{\mu}(k)+i0^{+}} \nonumber \\
  &&\times [f(E_{\nu}(k+q))-f(E_{\mu}(k))],\label{Eq:one}
\end{eqnarray}
where $\mu,\nu$ are the band indices and $q$ and $k$ are momentum vectors in the Brillouin zone and $N$ is the number of Fe lattice sites.

It is well-known that prominent condensation of particle-hole excitations at certain wave vector of $q$ in a disordered phase is a precursor for an appearance of corresponding symmetry breaking state as temperature becomes lower or interaction is switched on. In most of the iron-based superconductors, a remarkable condensation present at wave vector of $(\pi,\pi)$ can account for the stripe-type antiferromagnetic order or superconductivity. (Here we refer to a unit cell containing $2$ Fe atoms.) More detailedly, if the condensation at $(\pi,\pi)$ is strong enough, i.e., exceeds a threshold, stripe-type antiferromagnetic states tend to form. Otherwise tendency towards superconducting states mediated by short range antiferromagnetic fluctuations appears as long as the condensation around $(\pi,\pi)$ prevails over the others. Without noticeable condensation, disordered state remains.

\subsection{orbitally resolved momentum-dependent particle-hole excitations}
\label{sec:two}

In order to calculate the orbitally resolved momentum-dependent particle-hole excitations, an effective tight-binding model, including all the iron $d$ orbitals and the anion $p$ orbitals, are derived via construction of Wannier orbitals~\cite{wannier1,wannier2} on a 11$\times$11$\times$11 Monkhorst-Pack k-point mesh. The disentanglement procedure of wannier90 is employed in order to achieve the perfect match between the electronic structures of the effective tight-binding model and those obtained from {\it ab} initio calculations. Here the subspace selection step is done by projection and symmetry-preserved Wannier functions are adopted~\cite{wannierRMP}.

The condensation of the orbitally resolved momentum-dependent particle-hole excitations can be quantified by the real part of the static bare susceptibility of a multi-orbital system, defined as~\cite{graser}:
\begin{eqnarray}
  \chi^{pr;st}_{0} (q)&=& -\frac{1}{N} \sum_{k,\mu\nu}\frac{a^{s}_{\mu}(k)a^{p*}_{\mu}(k)a^{r}_{\nu}(k+q)a^{t*}_{\nu}(k+q)}{E_{\nu}(k+q)-E_{\mu}(k)+i0^{+}} \nonumber \\
  &&\times [f(E_{\nu}(k+q))-f(E_{\mu}(k))]\label{Eq:two}
\end{eqnarray}
where matrix elements $a^{s}_{\mu}(k)=\langle s|\mu k \rangle$ connect the orbital and the band spaces and are the components of the eigenvectors obtained from diagonalization of the effective tight-binding model. Here $f(E)$ is the Fermi distribution function, $p,r,s,t$ are the orbital indices, $\mu,\nu$ the band indices, $q$ and $k$ the momentum vectors in the Brillouin zone, and N the number of Fe lattice sites.

\section{Computational Results}

In iron-based superconductors, itinerant scenario states that Fermi surface nesting, i.e., particle-hole excitations among the Fermi surfaces condensed at $q=(\pi,\pi)$, determines the magnetism and superconductivity while localized picture emphasizes that the strong correlations are indispensable. In the following, we will show that the physical properties of various families of iron-based superconductors can be qualitatively understood in the absence of strong correlations, provides particle-hole excitations away from the Fermi level as well as orbital degrees of freedom and interatomic, interlayer coupling are properly taken into account.

\subsection{Comparison of Particle-hole Excitations close to and away from the Fermi Level}

\begin{figure}[htbp]
\includegraphics[width=0.48\textwidth]{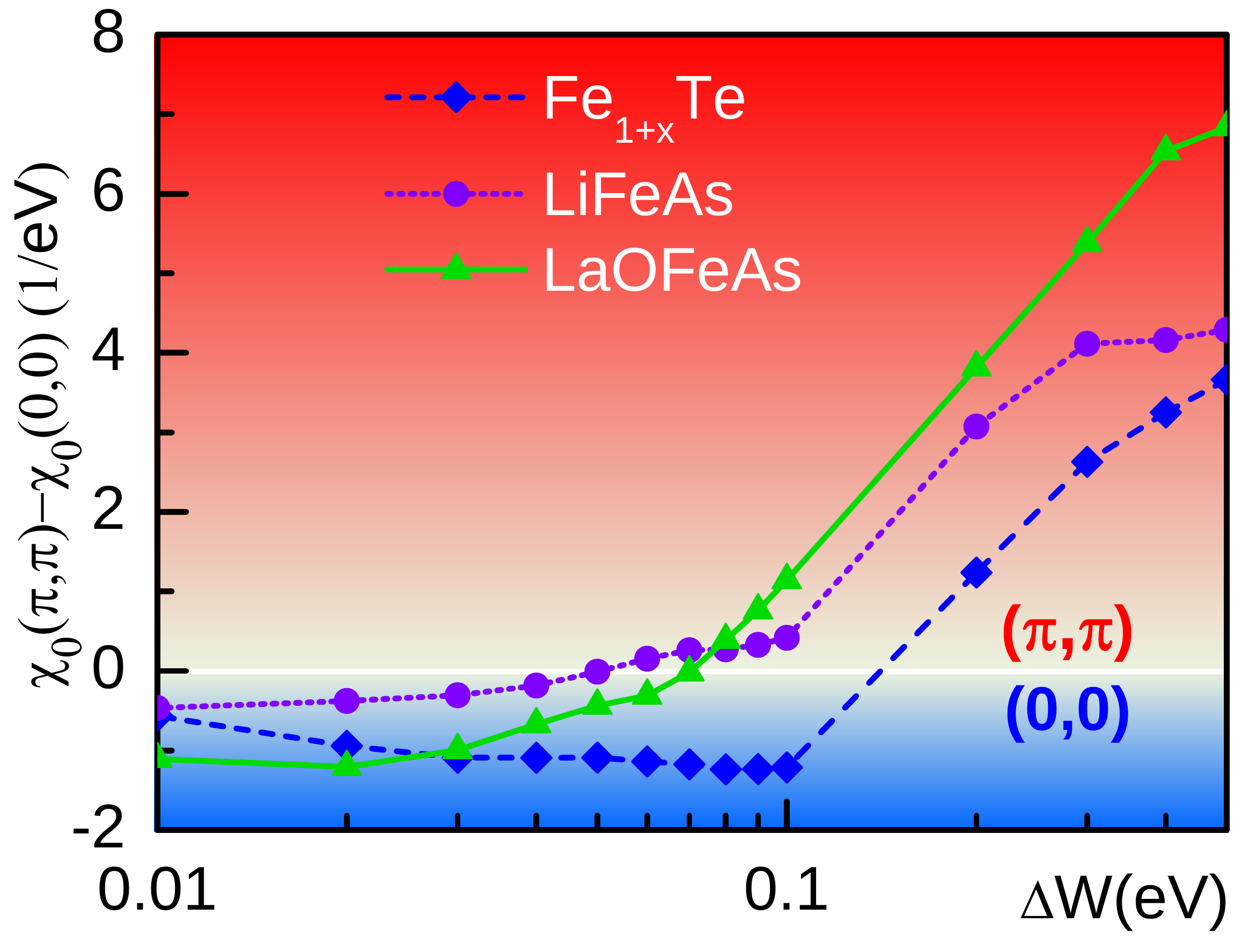}
\caption{The relative strength of particle-hole excitations calculated within the constant matrix elements approximation between $q=(\pi,\pi)$ and $(0,0)$ as a function of energy window chosen around the Fermi Level $E_F$ for the archetypal compounds FeTe, LiFeAs, and LaOFeAs. Here the energy window is defined as $[E_F-\Delta W,E_F+\Delta W]$. The condensation of particle-hole excitations is at $(0,0)$ for small energy window but is at $(\pi,\pi)$ when the energy window is large enough, indicating the importance of particle-hole excitations away from the Fermi level.}
\label{Fig:one}
\end{figure}

Fig.~\ref{Fig:one} shows the particle-hole excitations calculated within the constant matrix elements approximation~\cite{Mazin-Singh}~(see also Eq.~(\ref{Eq:one})) at wave vectors $(\pi,\pi)$ and $(0,0)$ for the representative compounds FeTe, LiFeAs, and LaOFeAs as a function of the width of energy window chosen around the Fermi Level, defined as $[E_F-\Delta W,E_F+\Delta W]$ where $E_F$ is the Fermi level and $2\Delta W$ is the energy window. Surprisingly, it is found that the particle-hole excitations are condensed at $(0,0)$ when the energy window is small. Only when the energy window is large enough are the particle-hole excitations condensed at $(\pi,\pi)$. This result suggests a new scenario that both the stripe-type antiferromagnetic states and the pairing for superconductivity mediated by spin fluctuations at $(\pi,\pi)$ are dominated by the electronic states away from the Fermi level, which is in stark contrast to the weak-coupling theory of Fermi surface nesting~\cite{Wanglee,Paglione,Mazin_PhysicaC,Mazin-Singh,Eremin,bkfeas,BaoFeTe,He-NaFeAs,DaiHu} where only the particle-hole excitations among the Fermi surfaces are emphasized. Though our results of large energy windows are consistent with previous studies, it is reported for the first time that the particle-hole excitations close to the Fermi level play an opposite role in deciding the magnetism and superconductivity.

\subsection{Condensed Particle-hole Excitations away from the Fermi Level at $(\pi,\pi)$}

Then, we will show that the new scenario is applicable to the whole family of iron-based compounds even if the strong correlation is completely absent. Since the nature of the condensation will be changed as a function of the energy window one chooses, throughout this paper we use a large enough energy window such that all the bands derived from the $d$ and $p$ orbitals are involved.

\begin{figure}[htbp]
\includegraphics[width=0.48\textwidth]{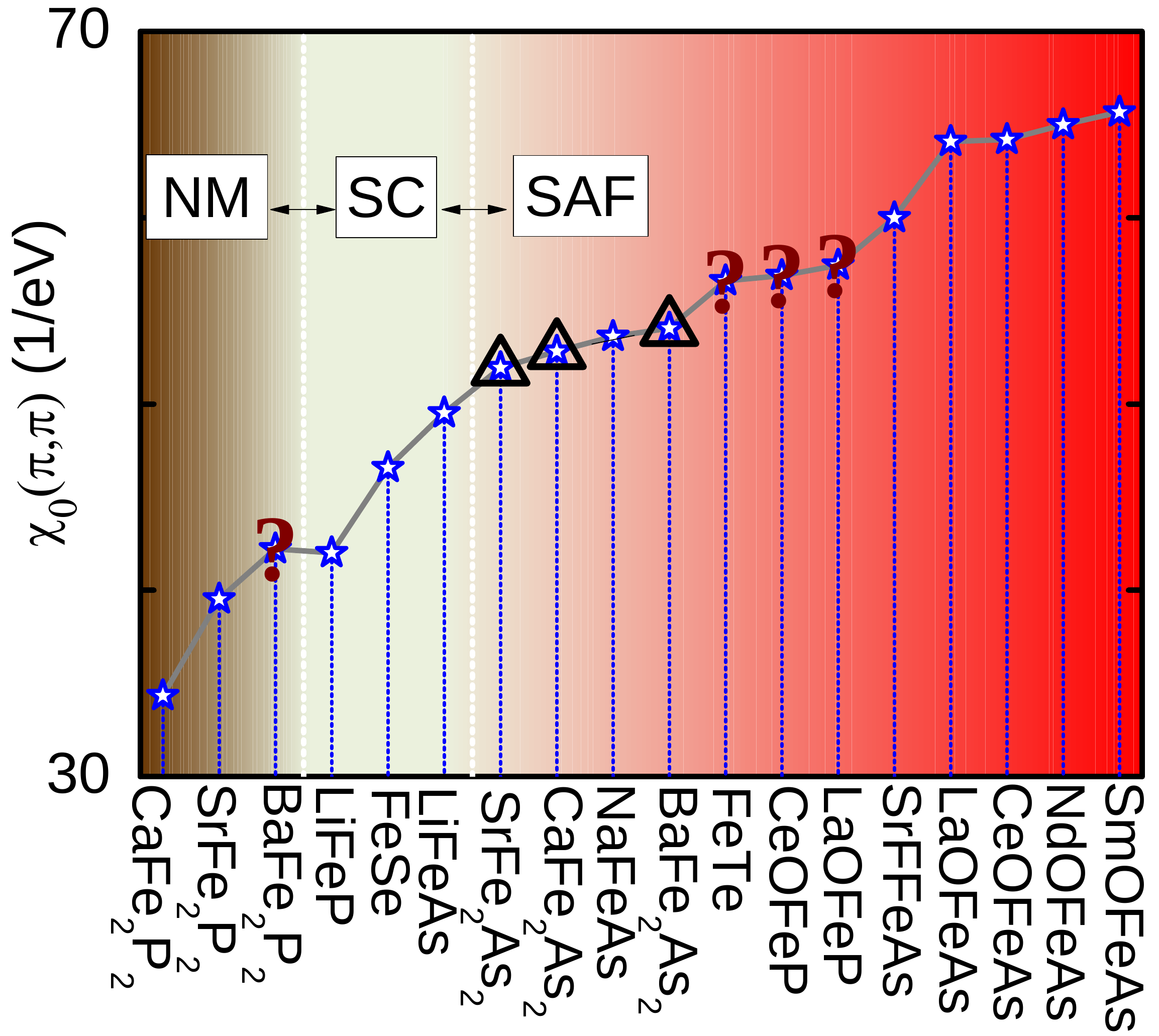}
\caption{Condensed particle-hole excitations within the constant matrix elements approximation at wave vector $(\pi,\pi)$ for various iron-based superconductors with large enough energy window. NM, SC, and SAF denote the nonmagnetic, superconducting, and stripe-type antiferromagnetic states, respectively. Most of the iron-based superconductors can be classified into the above three states by the strength of the condensation of particle-hole excitations at $(\pi,\pi)$ except for FeTe, LaOFeP, CeOFeP, and BaFe$_2$P$_2$, which are indicated by the question marks. The compounds AeFe$_2$As$_2$ with Ae=Ca, Sr, and Ba are marked by triangles due to the fact that their magnetic moments are larger than that of LaOFeAs, while the instabilities are weaker. The vertical dotted lines are a visual aid.}
\label{Fig:two}
\end{figure}

Fig.~\ref{Fig:two} shows the condensed particle-hole excitations at $(\pi,\pi)$ for various families of iron-based superconductors. Remarkably, it is found that the mechanism playing the dominant role in deciding the physical properties of the iron-based compounds is the condensed particle-hole excitations in momentum space, i.e., strong condensation of particle-hole excitations at $q=(\pi,\pi)$ leads to a stripe-type antiferromagnetic state while weak condensation results in a nonmagnetic phase. In the intermediate region, superconductivity appears.

\begin{figure*}[htbp]
\includegraphics[angle=-90,width=0.96\textwidth]{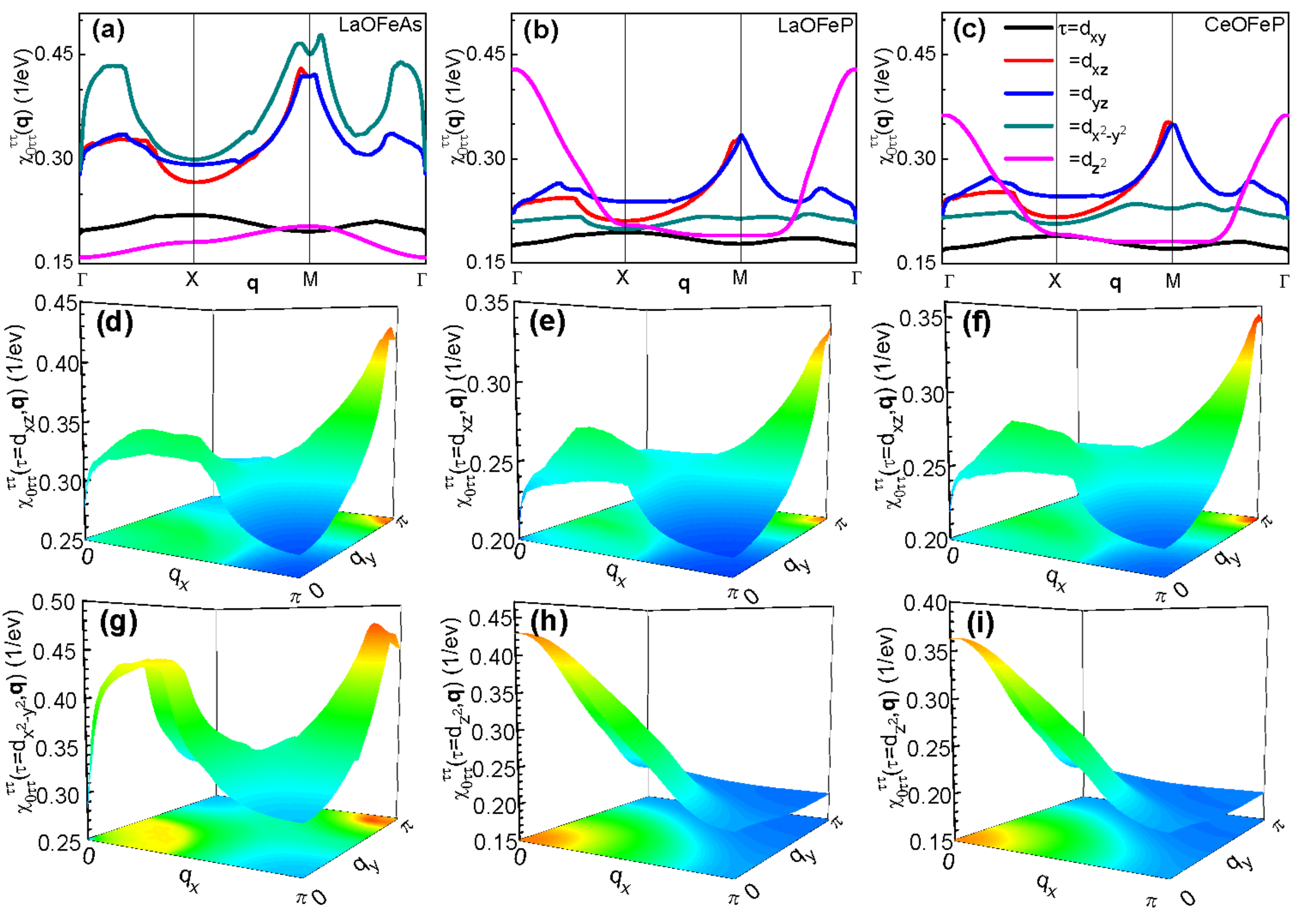}
\caption{Comparison of intra-orbital contributions to the particle-hole excitations among LaOFeAs, LaOFeP, and CeOFeP. (a), (b), (c) show five intra-orbital contributions along the path in momentum space from $\Gamma(0,0)$-$X(\pi,0)$-$M(\pi,\pi)$-$\Gamma(0,0)$ for LaOFeAs, LaOFeP, and CeOFeP, respectively. The dominating contributions from the d$_{xz}$ orbitals in the $q_x-q_y$ plane for LaOFeAs, LaOFeP, CeOFeP are presented in (d), (e), (f), respectively. The dominating contribution in the $q_x-q_y$ plane from the d$_{x^2-y^2}$ orbital in LaOFeAs and those from the d$_{z^2}$ orbitals in LaOFeP, CeOFeP are exhibited in (g), (h), (i), respectively. The two-dimensional contour maps are shown at the base of the figures.}
\label{Fig:three}
\end{figure*}

However, four compounds in Fig.~\ref{Fig:two} do violate the overall trend from this scenario and have been labelled with question marks. Specifically, LaOFeP and CeOFeP are superconducting~\cite{laofep} and heavy fermionic~\cite{ceofep} systems respectively while Fe$_{1+x}$Te is antiferromagnetically-ordered with a unique double stripe~\cite{BaoFeTe}, all of which are inconsistent with the strong condensation at $q=(\pi,\pi)$ in these materials. Moreover, BaFe$_2$P$_2$ is non-superconducting and nonmagnetic~\cite{bafepnesting,bafep}, but the condensation is slightly stronger than that of the superconductor LiFeP~\cite{lifep}. Furthermore, there is a quantitative problem present in Fig.~\ref{Fig:two} marked by three triangles which indicate that although the condensations are weaker in AeFe$_2$As$_2$ with Ae=Ca, Sr, Ba than in LaOFeAs, the observed magnetic moments are larger in AeFe$_2$As$_2$~\cite{laofeas,bafeas,cafeas,srfeas}. Nevertheless, as we will now show, all of the above anomalies can be simply explained.

\subsection{Multi-orbital effects}

We begin by explaining the qualitative anomalies. Whilst employing the constant matrix elements approximation does not qualitatively affect the results for the majority of the iron-based compounds, we find that it is necessary to drop this approximation and investigate the intra-orbital contributions to the particle-hole excitations~\cite{graser} (see also Eq.~(\ref{Eq:two})) in order to explain the qualitative anomalies. Such intra-orbital contributions to the particle-hole excitations have previously been found to play a dominant role in facilitating magnetism and superconductivity~\cite{Ding2013,leeshim,Kuroki}, compared to the inter-orbital counterparts. 

\subsubsection{LaOFeP: competitions among different orbitals}

First we will explain why strong condensation of particle-hole excitations at $(\pi,\pi)$ in superconducting LaOFeP as shown in Fig.~\ref{Fig:two} does not result in a stripe-type antiferromagnetic state. From Fig.~\ref{Fig:three} (a), (b), it is found that while the particle-hole excitations in the d$_{xz/yz}$ and d$_{x^2-y^2}$ orbitals are universally condensed around $(\pi,\pi)$ in the archetypal compound LaOFeAs indicating a single instability towards the stripe-type antiferromagnetic state, they are separately condensed at $(\pi,\pi)$ in the d$_{xz/yz}$ orbitals and at $(0,0)$ in d$_{z^2}$ orbitals in LaOFeP, suggesting competing tendencies towards either a stripe-type antiferromagnetic state or other magnetic states with $q=(0,0)$ such as Ne\'{e}l and ferromagnetically-ordered states. Here x, y, z are along the a, b, c directions, respectively, of the original unit cell with two iron atoms. Contributions from other orbitals are negligible compared to the above mentioned orbitals. Fig.~\ref{Fig:three} (d)-(i) further demonstrate that the particle-hole excitations are solely condensed either around $(\pi,\pi)$ or $(0,0)$ within the whole Brillouin zone in the dominating orbitals. It should be noted that the dominant contribution from the d$_{yz}$ orbital within the whole Brillouin zone is not shown since it can be reproduced by interchanging the $q_x$ and $q_y$ axes in Fig.~\ref{Fig:three} (d)-(f).

In order to reveal the competition between the tendencies towards different magnetically ordered states, we perform a mean-field calculation based on a multi-orbital Hubbard model where all the d orbitals from both Fe atoms in the unit cell are involved (see appendix~\ref{sec:app-two}). The tight-binding parameters are obtained through construction of Wannier orbitals as implemented in Wannier90~\cite{wannier1,wannier2}. A Hund's rule coupling of $J=0.15U$ is used for the calculations according to constrained random phase approximation calculations~\cite{CRPA}. 

\begin{figure}[htbp]
\includegraphics[width=0.48\textwidth]{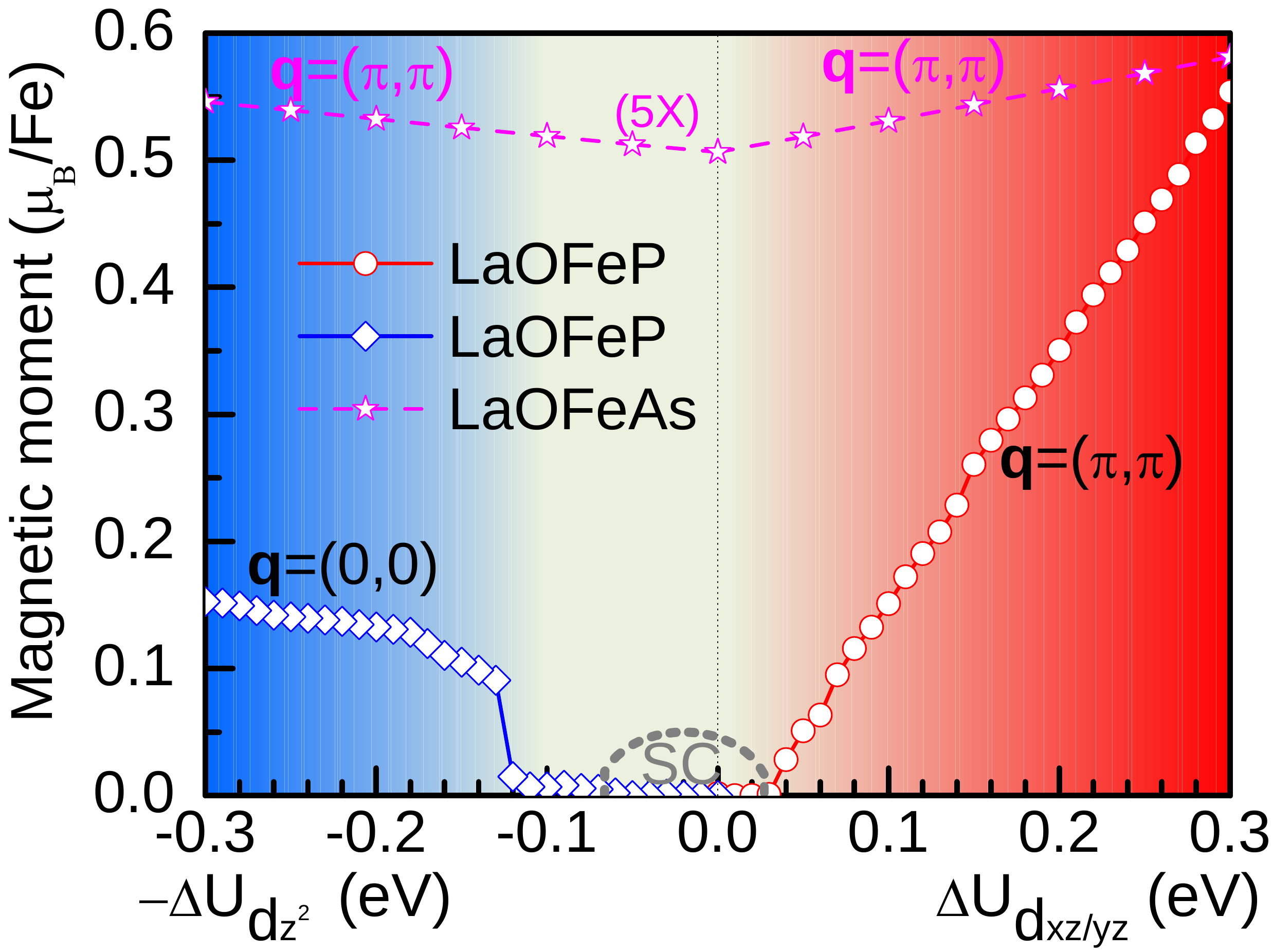}
\caption{Evolution of the ground state magnetic moments in LaOFeP and LaOFeAs as the intra-orbital Coulomb repulsion is slightly increased in the d$_{xz/yz}$ or d$_{z^2}$ orbital separately. Here a ten-orbital Hubbard model is constructed based on Wannier orbitals and $J/U=0.15$ is used according to the constrained random phase approximation calculations~\cite{CRPA}. Since our calculations are based on the mean-field approximation where quantum fluctuations are completely ignored, a comparatively smaller value of $U=1.7$~eV is used. It is shown that LaOFeP is located at a critical region as schematically indicated by a grey dome. The competition between the magnetically-ordered states with $q=(0,0)$ and $q=(\pi,\pi)$ remains if $U$ is changed.}
\label{Fig:four}
\end{figure}

Fig.~\ref{Fig:four} presents the evolution of the ground state magnetic moments in LaOFeP and LaOFeAs as the intra-orbital Coulomb repulsion in the d$_{xz/yz}$ and d$_{z^2}$ orbitals is varied individually. Since two competing condensations coexist in LaOFeP, it is expected that turning on the interaction and slightly increasing the intra-orbital Coulomb repulsion in the d$_{z^2}$ orbital alone will favor ferromagnetic or Ne\'{e}l ordered states according to the condensation at $q=(0,0)$. On the other hand, it is expected that the stripe-type antiferromagnetic state will be more strongly stabilized if the intra-orbital Coulomb repulsion in the d$_{xz/yz}$ orbitals are independently increased due to the condensation at $q=(\pi,\pi)$. This is indeed the case as shown in Fig.~\ref{Fig:four} for $U=1.7$~eV, indicating that LaOFeP is located in the proximity to the quantum critical region where magnetic orders with wave vectors $q=(0,0)$ and $q=(\pi,\pi)$ both vanish due to their mutual competition. Such competition can be viewed as an effective magnetic frustration due to itinerant electrons. The competition between the magnetically-ordered states with $q=(0,0)$ and $q=(\pi,\pi)$ remains if $U$ is changed. However, in the archetypal compound LaOFeAs, the stripe-type antiferromagnetic state always prevails over the other magnetic states since the dominating intra-orbital particle-hole excitations are all condensed at $(\pi,\pi)$.

\subsubsection{CeOFeP: effect of interatomic coupling}

Fig.~\ref{Fig:three} (b), (c) show that the momentum-dependent particle-hole excitations in CeOFeP is similar to those in LaOFeP when f electrons in Ce atoms are treated as nonmagnetic core electrons, implying that CeOFeP should also be a superconductor. However, CeOFeP is a heavy-fermion metal. This is due to the fact that the f electrons of the Ce atoms are not nonmagnetic core electrons and are strongly coupled with the d electrons of the Fe atoms, which suppress the antiferromagnetic fluctuations and consequently the tendency towards superconductivity. This can be revealed by comparisons of total energies among different magnetic states based on DFT calculations with an LDA+U functional applied to the f electrons of the Ce atoms. It is found that if the spins of the f electrons are unpolarized, the stripe-type antiferromagnetic state is the ground state. However, if the spins of f electrons are arranged ferromagnetically (as indicated experimentally~\cite{ceofep}) in the LDA+U calculations, the weakly-ferromagnetic solution has the lowest total energy, indicating that the couplings between f electrons of Ce atoms and d electrons of Fe atoms strongly affect the nature of the magnetic fluctuations in the Fe plane and therefore lead to a non-superconducting state.

\begin{figure}[htbp]
\includegraphics[width=0.48\textwidth]{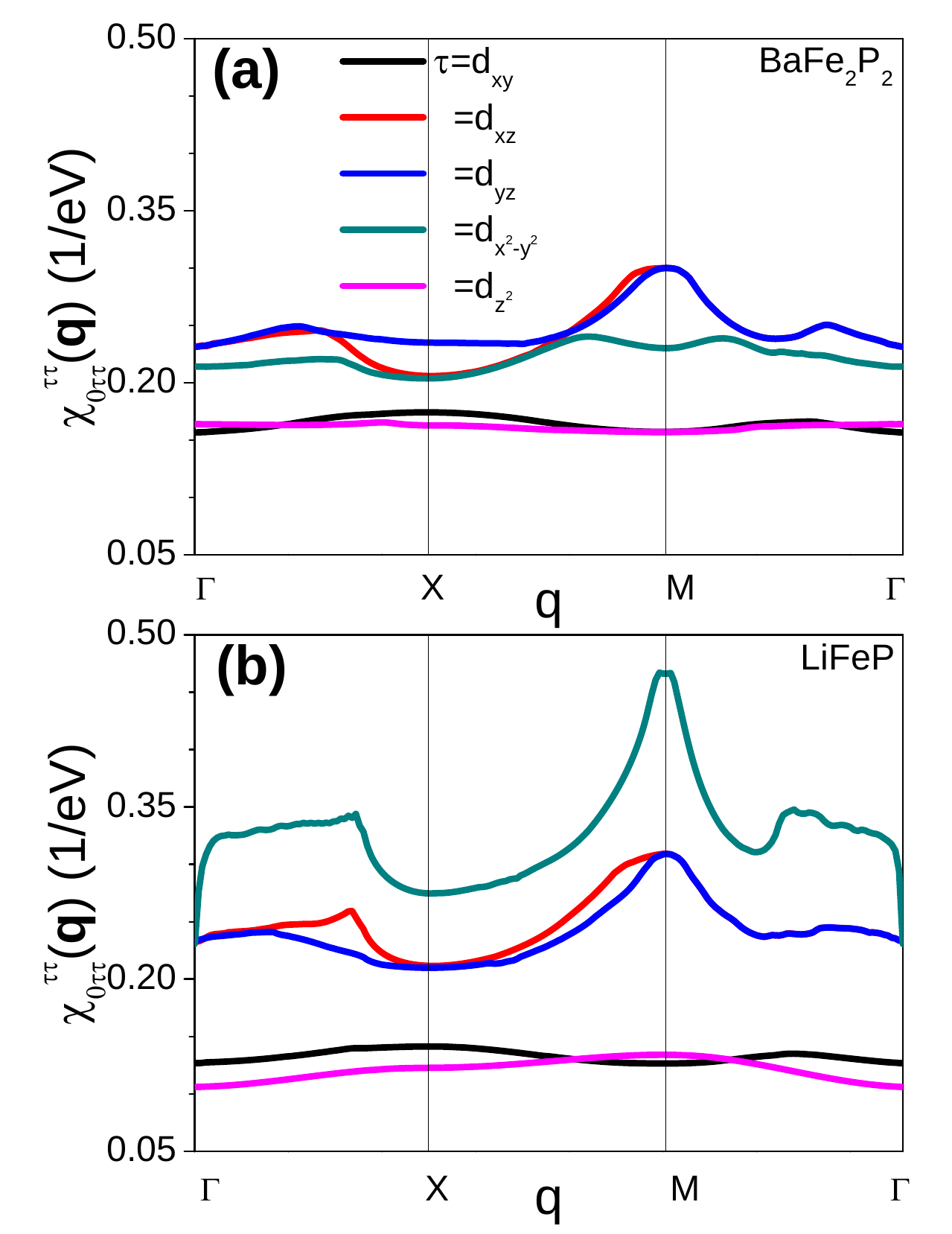}
\caption{Intra-orbital contributions to the particle-hole excitations. Five intra-orbital contributions along the path in momentum space from $\Gamma(0,0)$-$X(\pi,0)$-$M(\pi,\pi)$-$\Gamma(0,0)$ are shown for BaFe$_2$P$_2$ and LiFeP in (a) and (b), respectively.}
\label{Fig:five}
\end{figure}

\subsubsection{BaFe$_2$P$_2$: suppressed condensations of intra-orbital particle-hole excitations}

Next, we will show that inclusion of orbital degrees of freedom is enough to explain why BaFe$_2$P$_2$ is not a superconductor though the condensation of particle-hole excitations at $q=(\pi,\pi)$ calculated within the constant matrix elements approximation is stronger than that in the superconductor LiFeP. Fig.~\ref{Fig:five} shows the intra-orbital contributions to the particle-hole excitations in BaFe$_2$P$_2$ and LiFeP. In contrast to LiFeP, it is found that the condensation of particle-hole excitations in the d$_{x^2-y^2}$ orbital at $q=(\pi,\pi)$ vanishes in BaFe$_2$P$_2$. Only weak condensation in the d$_{xz/yz}$ orbitals remains at $q=(\pi,\pi)$, which is unlikely to be sufficient to support the appearance of either superconductivity or magnetism in BaFe$_2$P$_2$. Based on this scenario, we predict within the rigid-band approximation that K-doped BaFe$_2$P$_2$ can be a new candidate for an iron-based superconductor since the calculated intra-orbital contribution to the particle-hole excitations from the d$_{x^2-y^2}$ orbital at $q=(\pi,\pi)$ in hole-doped BaFe$_2$P$_2$ becomes as large as that in LiFeP when the hole concentration is around $0.25-0.3$~hole/Fe.

\begin{figure}[htbp]
\includegraphics[width=0.48\textwidth]{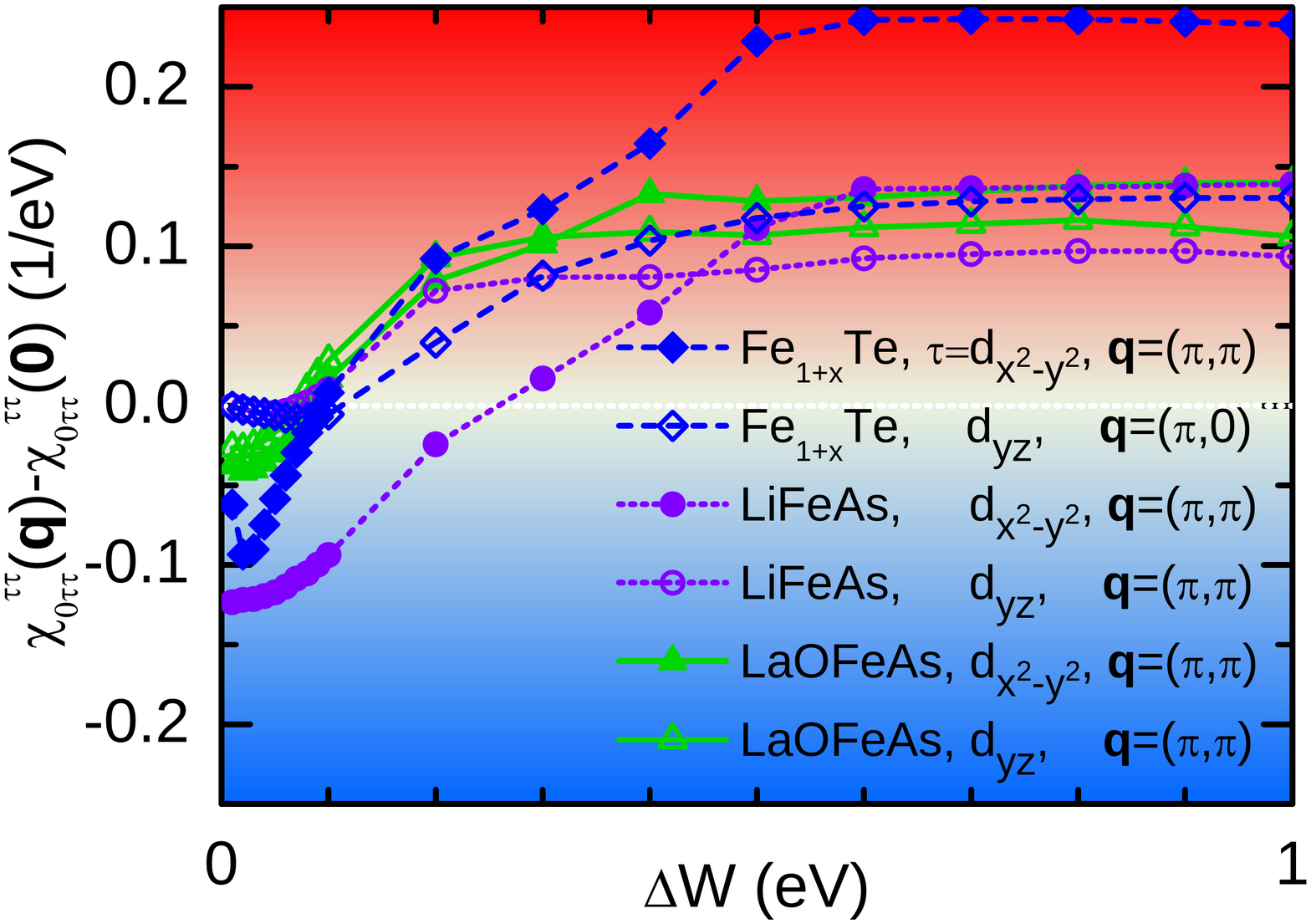}
\caption{The relative strength of intra-orbital contributions to the particle-hole excitations as a function of energy window chosen around the Fermi Level $E_F$ for FeTe, LiFeAs, and LaOFeAs. The particle-hole excitations become antiferromagnetic and the relative strength is saturated only when the energy window is large enough, $\Delta W > 200-600$~meV, indicating that electronic states away from the Fermi level play the dominant role in determining physical properties of iron-based superconductors.}
\label{Fig:six}
\end{figure}

\subsubsection{Multi-orbital effect on the relative strength of particle-hole excitations close to and away from the Fermi Level}

Then, it is interesting to investigate the effect of the multi-orbital physics on the nature of the condensed particle-hole excitations as a function of the width of the energy window chosen around the Fermi level. It is shown in Fig.~\ref{Fig:six} that in all the compounds we studied, the relative magnitudes of the condensation become saturated only when the energy window is large enough, at around $200-600$~meV. However, in the vicinity of Fermi level, i.e., with very small energy window or when the energy window less than $50$~meV, the particle-hole excitations in all cases are condensed at $(0,0)$ rather than antiferromagnetically at $(\pi,\pi)$ or $(\pi,0)$, which is qualitatively consistent with the results from the constant matrix elements approximation as shown in Fig.~\ref{Fig:one}.

\subsubsection{Fe$_{1+x}$Te: effect of excess Fe}

Moreover, a proposal has already been made to understand the unique double-stripe antiferromagnetism in Fe$_{1+x}$Te based on the scenario of condensed particle-hole excitations~\cite{Ding2013}. The role of excess Fe in the interstitial was emphasized since it contributes not only excess electrons to the in-plane Fe which enhances the tendency towards the double-stripe antiferromagnetic state but also a magnetic ion strongly coupled with the in-plane Fe~\cite{Ding2013} which further suppresses other magnetic instabilities~\cite{Liunmat}. Please note, this proposal is valid irrespective of the approximations one chooses, while other theories strongly depend on the approximations, i.e., LDA and GGA lead to contradictory conclusions~\cite{Ding2013}.

\begin{table}
  \caption{Inter-layer exchange couplings of various iron-based superconductors. Here inter-layer exchange couplings $J=\Delta E= E_{SAF,II}-E_{SAF,I}$, where (SAF,II) denotes the intra-layer stripe-type antiferromagnetic state with inter-layer antiferromagnetic spin arrangement and (SAF,I) refers to the intra-layer stripe-type antiferromagnetic states with inter-layer ferromagnetic spin arrangement, are calculated within both the local density approximation (LDA) and the generalized gradient approximation (GGA) with units meV/Fe.}
\label{Tab:one}
\begin{ruledtabular}
\begin{tabular}{@{}ccccccc@{}}
& $\Delta E_{LDA}$ & $\Delta E_{GGA}$ \\\hline\hline
 CaFe$_2$As$_2$   &  -25.19   &  -32.55  \\
 SrFe$_2$As$_2$   &  -10.12  &  -13.03  \\
 BaFe$_2$As$_2$  &  -2.90  &  -4.39  \\
 NaFeAs   &  -0.72   &  -0.91  \\
 LaOFeAs   &  -0.02   &  -0.04  \\
\end{tabular}
\end{ruledtabular}
\end{table}

\subsection{Importance of interlayer coupling}

Finally, we explain the reason behind the quantitative anomalies illustrated in Fig.~\ref{Fig:two}. We emphasize that it is the inter-layer couplings ignored in most cases that are responsible for the quantitative inconsistency between weaker condensations but larger magnetic moments in AeFe$_2$As$_2$ with Ae=Ca, Sr, Ba, in comparison to LaOFeAs. We point out that although all the iron-based superconductors are layered materials, the band structures of AeFe$_2$As$_2$ compounds are more three-dimensional than that of LaOFeAs. As is well-known, the higher the dimensionality, the lower the fluctuations suppressing the ordered state. This reasoning can be further corroborated by investigating the strength of inter-layer exchange couplings in different families of iron-based superconductors. As shown in Table~\ref{Tab:one}, the inter-layer couplings in AeFe$_2$As$_2$ compounds are at least one order of magnitude larger than the other families such as the [111] and [1111] compounds. Therefore although the condensation is weaker in AeFe$_2$As$_2$ than in LaOFeAs, the stronger inter-layer couplings can suppress the fluctuations and hence enhance the magnetically-ordered state.

\section{Conclusion}

The origin of magnetism and superconductivity in all families of iron-based superconductors, except KFe$_2$Se$_2$ whose parent compound is still under debate, can be qualitatively understood from the scenario of condensed particle-hole excitations away but not far from the Fermi level in the absence of strong correlations, indicating that the strong correlation may not play a crucial role in determining the physics of iron-based superconductors, which is in contrast to the high-T$_c$ cuprates. However, the scenario has no relation with the conventional weak-coupling theory of Fermi surface nesting. The particle-hole excitations within $E_F \pm 200 \sim 600$~meV, rather than the Fermi surface or the one-particle states below the Fermi level, determine the physical properties of iron-based superconductors. The orbital degrees of freedom and interlayer couplings have to be involved in order to correctly understand some anomalies in iron-based compounds. The competing tendencies towards different magnetic states coexisting in different orbitals are the itinerant analogy to the magnetic frustration and the proximity to an antiferromagnetic quantum critical point are responsible for the superconductivity in the iron-based superconductors with nonmagnetic parent states, such as LaOFeP. Our results have a broad implication that a new theory, rather than an extension of conventional single-band theory, is required if one wants to correctly understand real materials with multiple active bands crossing the Fermi level.

\section{Acknowledgement}

This work is supported by National Natural Science Foundation of China (No. 11174219), Program for New Century Excellent Talents in University (NCET-13-0428), Research Fund for the Doctoral Program of Higher Education of China (No. 20110072110044) and the Program for Professor of Special Appointment (Eastern Scholar) at Shanghai Institutions of Higher Learning as well as the Scientific Research Foundation for the Returned Overseas Chinese Scholars, State Education Ministry.

\appendix

\section{Experimental structures for all the compounds we studied}
\label{sec:app-one}
The experimental structures we used in this study are shown in Table~\ref{Tab:app-one}.

\section{Multi-orbital Hubbard model}
\label{sec:app-two}

The ten-orbital Hubbard model is defined as:
\begin{eqnarray}
&H&=-\sum_{ ij, \gamma \gamma' \sigma } t_{ij,\gamma \gamma'} c^{\dagger}_{i\gamma\sigma}c_{j\gamma'\sigma}
+U\sum_{i\gamma}n_{i\gamma\uparrow}n_{i\gamma\downarrow} \nonumber \\
&+&\Big(U'-\frac J 2\Big)\sum_{i\gamma>\gamma'}n_{i\gamma} n_{i\gamma'}-2J\sum_{i\gamma>\gamma'}S_{i\gamma}^{z} S_{i\gamma'}^{z} \nonumber \\
&+&\sum_{i,\gamma=d_{xz},d_{yz},d_{z^2}} \Delta U_{\gamma}n_{i\gamma\uparrow}n_{i\gamma\downarrow},\label{eq:hamiltonian}
\end{eqnarray}
where $t_{ij,\gamma\gamma'}$ are hopping integrals between different sites $i$ and $j$ with orbital indices $\gamma,\gamma'$ ranging from $1$ to $10$ which represent the $d_{xy}, d_{xz}, d_{yz},d_{x^2-y^2}, d_{z^2}$ orbitals of both Fe atoms in the unit cell. $U$, $U^{\prime }$ and $J$ are the intra-orbital, inter-orbital Coulomb interaction and Hund's rule coupling, respectively, which fulfill the rotational invariance condition $U=U^{\prime }+2J$. The pair-hopping and spin-flip terms are ignored as those do not affect our mean-field results qualitatively ~\cite{Lorenzana2008,Daghofer2008,Kaneshita2009,Bascones2010}.
$c^{\dagger}_{i\gamma\sigma}$ ($c_{i\gamma\sigma}$) creates (annihilates) an electron in orbital $\gamma$ of site $i$ with spin $\sigma$. $n_{i\gamma\sigma}$ is the occupation operator, $n_{i\gamma}=n_{i\gamma\uparrow}+n_{i\gamma\downarrow}$, and $S_{i\gamma}^{z}$ the spin operator in $z$ axis. The around-mean-field double counting correction is taken into account.

The last term is used to adjust separately the strength of the intra-orbital Coulomb repulsions in d$_{xz/yz}$ and d$_{z^2}$ orbitals, which can be applied to manifest the competition of the tendencies towards different magnetic states decided by d$_{xz/yz}$ and d$_{z^2}$ orbitals. It is expected that in LaOFeP, since two competing condensations coexist, slightly increasing the intra-orbital Coulomb repulsion ($\Delta U_{d_{z^2}}$) in d$_{z^2}$ orbital alone will favor ferromagnetic or Ne\'{e}l ordered state according to the condensation at $q=(0,0)$ while the stripe-type antiferromagnetic state will be strongly stabilized if the intra-orbital Coulomb repulsions ($\Delta U_{d_{xz/yz}}$) in d$_{xz/yz}$ orbitals become a bit larger independently, due to the condensation at $q=(\pi,\pi)$.

\begin{table}
  \caption{The lattice constants for all the compounds we studied.}
\label{Tab:app-one}
\begin{ruledtabular}
\begin{tabular}{@{}ccccccc@{}}
& a(b)(\AA) & c(\AA) \\\hline\hline
 SmOFeAs~\cite{SmOFeAs-str}   &  3.9391   &  8.4970  \\
 NdOFeAs~\cite{NdOFeAs-str}   &  3.9611  &  8.5724  \\
 CeOFeAs~\cite{CeOFeAs-str}  &  3.99591  &  8.6522  \\
 LaOFeAs~\cite{LaOFeAs-str}   &  4.03533   &  8.74090  \\
 SrFFeAs~\cite{SrFFeAs-str}   &  3.9996   &  8.9618  \\
 LaOFeP~\cite{laofep}   &  3.96358   &  8.51222  \\
 CeOFeP~\cite{ceofep,ceofep-str2}   &  3.9195   &  8.3273  \\
 FeTe~\cite{FeTe-str}   &  3.82134   &  6.2517  \\
 BaFe$_2$As$_2$~\cite{bafeas} &  3.95702   &  12.9685  \\
 NaFeAs~\cite{NaFeAs-str}   &  3.94481  &  6.9968  \\
 CaFe$_2$As$_2$~\cite{cafeas-str} & 3.8915  &  11.69  \\
 SrFe$_2$As$_2$~\cite{srfeas-str}   &  3.39243   &  12.3644  \\
 LiFeAs~\cite{LiFeAs-str}  &  3.7914   &  6.3639  \\
 FeSe~\cite{FeSe-str}   &  3.7727   &  5.526  \\
 LiFeP~\cite{lifep}   &  3.69239   &  6.03081  \\
 BaFe$_2$P$_2$~\cite{bafep-str}  & 3.840  &  12.4420 \\
 SrFe$_2$P$_2$~\cite{bafep-str}   &  3.8250   &  11.6120  \\
 CaFe$_2$P$_2$~\cite{bafep-str}   & 3.8550   &  9.9850  \\
\end{tabular}
\end{ruledtabular}
\end{table}

\end{document}